\documentclass[twocolumn,prl,aps,showpacs]{revtex4}
\usepackage{graphicx,amsmath,amssymb}

\newcommand{\abbrev}{\small}

\newcommand{\eqn}[1]{Eq.\,(\ref{#1})}

\newcommand{\reference}[1]{Ref.\,\cite{#1}}
\newcommand{\refs}[1]{Refs.\,\cite{#1}}
\newcommand{\fig}[1]{Fig.\,(\ref{#1})}

\newcommand{\api}{\frac{\alpha_s}{\pi}}
\newcommand{\nnlo}{{\abbrev NNLO}}
\newcommand{\qcd}{{\abbrev QCD}}
\newcommand{\msbar}{\overline{\mbox{\small MS}}}

\newcommand{\logmut}{l_{\mu t}}

\newcommand{\mhiggs}{M_H}
\newcommand{\lhc}{{\abbrev LHC}}

\newcommand{\lnxm}[1]{L^{#1}(x)}
\newcommand{\Li}[1]{{\mathop{\rm Li}_{#1}\nolimits}}
\newcommand{\ARcite}[2]{{[\citealp{#1}, \citealp{#2}]}}

\begin{document}

\begin{titlepage}

\hspace*{\fill}\parbox[t]{4cm}{
BNL-HET-02/3\\
CERN-TH/2002-006\\
hep-ph/0201206\\
January 2002}

\bibliographystyle{apsrev}
\preprint{BNL-HET-02/3}
\preprint{CERN-TH/2002-006}
\title{ Next-to-Next-to-Leading Order Higgs Production at Hadron Colliders }
\author{Robert~V.~Harlander}
\affiliation{  TH Division, CERN, CH-1211 Geneva 23, Switzerland\\
  {\tt [robert.harlander@cern.ch]}}
\author{William~B.~Kilgore}
\affiliation{Physics Department, Brookhaven National Laboratory,
  Upton, New York
  11973, U.S.A.\\
  {\tt [kilgore@bnl.gov]} }
\begin{abstract}
  The Higgs boson production cross section at $pp$ and $p\bar p$
  colliders is calculated in \qcd\ at next-to-next-to-leading order
  (\nnlo).  We find that the perturbative expansion of the production
  cross section is well behaved and that scale dependence is reduced
  relative to the {\abbrev NLO} result.  These findings give us
  confidence in the reliability of the prediction.  We also report an
  error in the {\abbrev NNLO} correction to Drell-Yan production.
\end{abstract}
\maketitle
\end{titlepage}


\section{Introduction}
Gluon fusion will be the most important production channel for Higgs
discovery at the \lhc. The Higgs boson should manifest itself in the
reaction $pp\to H(\to\gamma\gamma) + X$, where a signal should emerge on
top of a very smooth, measurable $\gamma\gamma$ background.
At the Tevatron, the focus for Higgs discovery is in associated
production modes like $W/Z+H$ and $t\bar t + H$. In a mass window
around the $WW$ threshold, however, gluon fusion is important.

\nnlo{} corrections to the process $gg\to H$ have been evaluated
recently in the heavy top limit and the approximation of soft gluon
radiation \cite{Harlander:2000mg,Harlander:2001is,Catani:2001ic},
where the partonic center-of-mass energy is close to the Higgs mass,
$\mhiggs^2/\hat{s} \equiv x \to 1$.  If we write the partonic cross
section as an expansion in $(1-x)$, it has the following form:
\begin{equation}
\begin{split}
  \hat\sigma_{ij} &= \sum_{n\geq 0}
  \left(\api\right)^n\,\hat\sigma_{ij}^{(n)}\,,\\
  \hat\sigma_{ij}^{(n)} &= a^{(n)}\,\delta(1-x) +
  \sum_{k=0}^{2n-1}b_k^{(n)}\left[\frac{\ln^k(1-x)}{1-x}\right]_+ \\
  &\quad+ \sum_{l=0}^\infty\sum_{k=0}^{2n-1}
  c_{lk}^{(n)}\, (1-x)^l\ln^k(1-x)\,,
\label{eq::abc}
\end{split}
\end{equation}
where the $[\ ]_+$ terms are ``$+$''-distributions defined in the
usual way (see, e.g.\,\reference{Harlander:2001is}).
\refs{Harlander:2000mg,Harlander:2001is,Catani:2001ic} contain the
coefficients $a^{(n)}$ and $b_k^{(n)}$ up to $n=2$ of this expansion.
However, as anticipated in \reference{Kramer:1998iq}, these
contributions are not sufficient to arrive at a reliable prediction
for the total cross section.  Using resummation techniques, the
authors of \reference{Kramer:1998iq} evaluated the coefficient
$c_{03}^{(2)}$ at \nnlo{}. It was included in the final results of
\refs{Harlander:2001is,Catani:2001ic}.\break  However, the unknown
sub-leading terms $c_{0i}^{(2)}$ with $i\le 2$, were treated in
different ways by \refs{Harlander:2001is} and \cite{Catani:2001ic},
leading to significant deviations in the numerical results.  It is the
purpose of the current letter to report on the analytical evaluation
of the coefficients $c_{lk}^{(2)}$ with $k=0,\ldots,3$ and $l\geq
0$. In other words, we compute the partonic cross section for Higgs
production in terms of an expansion around the soft limit. We find
that the series converges very well and conclude that our final
results are equivalent to a calculation of the cross section in closed
analytic form.  We therefore resolve the ambiguities of
\refs{Harlander:2001is,Catani:2001ic} and provide a realistic
prediction for the Higgs production cross section in $pp$ and $p\bar
p$ collisions.

In checking our methods, we found an error in the\break \nnlo{} Drell-Yan
calculation of \reference{Hamberg:1991np}. The correct result is
given at the end of the next section.


\section{The Calculation}
In the following we will assume all quark masses to vanish, except for
the top quark mass, and neglect all electro-weak couplings. In this
limit, the Higgs boson can couple to gluons only via a top quark loop.
This coupling can be approximated by an effective Lagrangian
corresponding to the limit $m_t\to \infty$, which is valid for a large
range of $\mhiggs$, including the currently favored region between 100
and 200\ GeV.  The effective Lagrangian is
\begin{equation}
\begin{split}
\label{eq::efflag}
{\cal L}_{\rm eff} &= -\frac{H}{4v}C_1(\alpha_s)\,G_{\mu\nu}^aG^{a\,\mu\nu}\,,
\end{split}
\end{equation}
where $G_{\mu\nu}^a$ is the gluon field strength tensor, $H$ is the
Higgs field, $v\approx 246$\,GeV is the vacuum expectation value of
the Higgs field and $C_1(\alpha_s)$ is the Wilson coefficient.
Renormalization of this Lagrangian has been discussed in
\reference{Harlander:2001is}, for example, and shall not be repeated
here. In the $\msbar$ scheme, the coefficient function $C_1(\alpha_s)$
reads, up to the order required here
\ARcite{Chetyrkin:1997iv}{Kramer:1998iq}:
\begin{equation}
\begin{split}
C_1(\alpha_s) &= -\frac{1}{3}\api\bigg\{
1 + \frac{11}{4}\api
+ \left(\api\right)^2\bigg[\frac{2777}{ 288} 
\\&\qquad
+ \frac{19}{ 16}\logmut{}
+ n_f\left(-\frac{67}{ 96} + \frac{1}{ 3}\logmut{}\right)\bigg]
+ \ldots\bigg\}\,,
\label{eq::c1}
\end{split}
\end{equation}
where $\logmut{} = \ln(\mu_R^2/M_t^2)$. $\mu_R$ is the renormalization scale
and $M_t$ is the on-shell top quark mass.  $\alpha_s \equiv
\alpha_s^{(5)}(\mu_R^2)$ is the $\msbar$ renormalized \qcd\ coupling
constant for five active flavors, and $n_f$ is the number of massless
flavors.  In our numerical results, we always set $n_f = 5$.

The Feynman diagrams to be evaluated for hadronic collisions
at \nnlo{} are: {\rm (i)}~two-loop virtual diagrams for $gg\to H$;
{\rm (ii)}~one-loop single real emission diagrams for $gg\to Hg$,
$gq\to Hq$, and $q\bar q\to Hg$; {\rm (iii)}~tree-level double real
emission diagrams for $gg\to Hgg$, $gg\to Hq\bar q$, $gq\to Hgq$,
$qq\to Hqq$, $q\bar q\to Hgg$, and $q\bar q\to Hq\bar q$.  The
coefficients $a^{(2)}$ and $b^{(2)}_k$ in \eqn{eq::abc} are determined
by the $gg$ sub-process only, while the $c_{lk}^{(2)}$ receive
contributions from all sub-processes.

For the single real emission diagrams {\rm (ii)}, the full analytical
result for general values of $x$ has been evaluated and will be
published elsewhere.  It can be expanded trivially in terms of
$(1-x)$.  In order to obtain this expansion for the double real
emission contribution {\rm (iii)}, we evaluated the squared amplitude
and expressed the invariants of incoming and outgoing momenta, as well
as the phase space measure in terms of two scattering angles and the
dimensionless variables $x$, $y$, $z$, defined
by~\cite{Matsuura:1989sm}
\begin{equation}
\begin{split}
\mhiggs^2 &= \hat{s}\,x\,,\qquad
p_1\cdot p_H = \frac{\hat{s}}{ 2}(1-(1-x)y)\,,\\
p_2\cdot p_H &= \frac{\hat{s}}{ 2}\left( \frac{x + (1-x)^2y(1-y)(1-z)}{
    1-(1-x)y} \right)\,.
\end{split}
\end{equation}
$p_1$, $p_2$, and $p_H$ are the momenta of the incoming partons\break
and the (outgoing) Higgs boson, respectively.  Modulo powers that
vanish as $d\to 4$ ($d$ is the space-time dimension), the resulting
expression is expanded as a Laurent series in $(1-x)$. The leading
terms in this expansion are of order $(1-x)^{-1}$ and give rise to the
purely soft contribution obtained in
\refs{Harlander:2001is,Catani:2001ic}. Here we also keep higher orders
in this expansion, $(1-x)^l$, $l\geq 0$.  Aside from a few extra
algebraic manipulations of hypergeometric functions, this expansion
procedure allows us to perform\break the phase space integration along the
lines of \reference{Matsuura:1989sm}. Details of the calculations will be
presented elsewhere.

There are a number of checks that can be performed on our result.  One
is to see that all poles in $d-4$ cancel.  We have explicitly verified
this cancellation through order\break $(1-x)^{16}$.  Since we have
computed single real emission and the mass factorization counterterms
in closed form\break (as opposed to an expansion in $(1-x)$), we can
also obtain the pole terms for double real emission in closed form by
{\it demanding} that the poles cancel.  This allows us to obtain in
closed form all finite terms in the cross section that are linked to
the poles.  These include all terms proportional to $\ln^n(1-x)$
($n=1,2,3$) and all explicitly scale-dependent terms.

As another check on our approach we applied it to the cross section
for the Drell-Yan process at \nnlo{}, where the full $x$-dependence is
known in analytical form~\cite{Hamberg:1991np}.  Detailed comparison
with unpublished intermediate results~\footnote{We thank
W.L.~van~Neerven and V.~Ravindran for confirming these corrections to
the Drell-Yan cross section.} 
shows that our expansion of the
tree-level double real emission terms is in complete agreement with
the corresponding expansion of the exact calculation.\break However,
we find differences in the one-loop single real emission terms which
we also have computed exactly.  We conclude that the \nnlo{} result
for the Drell-Yan process in \reference{Hamberg:1991np} is incorrect
and that the correct result is
\begin{equation}
\begin{split}
\label{eq::dycorrect}
\Delta^{(2),C_A}_{q\bar{q}} &=
    \left.\Delta^{(2),C_A}_{q\bar{q}}
    \right|_{\mbox{\scriptsize\reference{Hamberg:1991np}}}
    + \left(\frac{\alpha_s}{4\pi}\right)^2 C_A C_F\,\Big\{\\
  -&\,8x\left(2\Li2(1-x) + 2\ln(x)\ln(1-x) - \ln^2(x)\right)\Big\}\,,\\
\Delta^{(2),C_F}_{q\bar{q}} &=
    \left.\Delta^{(2),C_F}_{q\bar{q}}
    \right|_{\mbox{\scriptsize\reference{Hamberg:1991np}}}
    + \left(\frac{\alpha_s}{4\pi}\right)^2 C^2_F\,\Bigl\{
      -16\ln(x)\\
  -&\,8(3+x)\left(2\Li2(1-x) + 2\ln(x)\ln(1-x) - \ln^2(x)\right)
      \Bigr\}\,,\\
\Delta^{(2),C_A}_{qg} &=
    \left.\Delta^{(2),C_A}_{qg}
    \right|_{\mbox{\scriptsize\reference{Hamberg:1991np}}}
    + \left(\frac{\alpha_s}{4\pi}\right)^2 C_A T_f\Bigl\{
      - 8x\ln(x)\\
  +&\,4x\left(2\Li2(1-x) + 2\ln(x)\ln(1-x) - \ln^2(x)\right)
      \Bigr\}\,,\\
\Delta^{(2),C_F}_{qg} &=
    \left.\Delta^{(2),C_F}_{qg}
    \right|_{\mbox{\scriptsize\reference{Hamberg:1991np}}}
    + \left(\frac{\alpha_s}{4\pi}\right)^2 C_F T_f\Bigl\{
      \vphantom{\left(\ln^2(x)\right)}\\
  -&\,4(3-x)\left(2\Li2(1-x) + 2\ln(x)\ln(1-x) - \ln^2(x)\right)\\
  +&\,12(1-x)(1 - 2\ln(1-x)) + (28 - 44x)\ln(x)
      \vphantom{\left(\ln^2(x)\right)}\Bigr\}\,.
\end{split}
\end{equation}
The numerical effect of these corrections is rather small and shall be
investigated in more detail elsewhere.


\section{Partonic Results}
We now present the result for the partonic Higgs production cross
sections at \nnlo{}.  We define
\begin{equation}
\begin{split}
\sigma_0 &= \frac{\pi}{ 576v^2}\left(\api\right)^2\,,
\qquad\lnxm{} = \ln(1-x)\,,\\
\hat\sigma_{ij} &= \hat\sigma^{(0)}_{ij} + \api\,\hat\sigma^{(1)}_{ij} +
\left(\api\right)^2\,\hat\sigma^{(2)}_{ij} + \ldots\,.
\end{split}
\end{equation}
The lower order terms, $\hat\sigma^{(0)}_{ij}$ and
$\hat\sigma^{(1)}_{ij}$, are given in\break
\refs{Dawson:1991zj,Djouadi:1991tk}.  If we split the second order
terms into ``soft'' and ``hard'' pieces,
\begin{equation}
\begin{split}
\hat\sigma_{ij}^{(2)} &= \delta_{ig}\delta_{jg}\, \hat\sigma_{gg}^{
  (2),\rm soft} + \hat\sigma_{ij}^{(2),\rm h}\,,
\end{split}
\end{equation}
the soft pieces are given in Eq.\,(25) of
\reference{Harlander:2001is},
while the hard pieces, $\hat\sigma_{ij}^{(n),\rm h}$ (to order
$(1-x)^1$) are:
\begin{equation}
\begin{split}
&\hat\sigma_{gg}^{(2),\rm h} =
  \sigma_0
\bigg\{
 \frac{1453}{12} 
  - 147\,\zeta_2 
  - 351\,\zeta_3 
  + n_f\bigg( 
    - \frac{77}{18} 
    + 4\,\zeta_2 
  \bigg)  
\\&\mbox{}
+ \lnxm{}\,\bigg[
  - \frac{1193}{4} 
  + 180\,\zeta_2 
  + \frac{101}{12} n_f
\bigg]
\\&\mbox{}
+ \lnxm{2}\left( 
  \frac{411}{2} 
  - 4\,n_f
\right)
- 144\,\lnxm{3}
\\&
+ (1-x)\bigg[
-\frac{3437}{4} 
+ \lnxm{}\left(\frac{2379}{2} - 270\,\zeta_2\right) 
\\&
- \frac{2385}{4}\,\lnxm{2} + 216\,\lnxm{3} 
+ \frac{1017}{2}\,\zeta_2 
+ \frac{1053}{2}\,\zeta_3
\\&
+ n_f\,\bigg(
\frac{395}{24} - \frac{45}{2}\,\lnxm{} + \frac{22}{3}\,\lnxm{2} 
- \frac{22}{3}\,\zeta_2\bigg)\bigg]
+\ldots
\bigg\}\,,
\label{eq::ggxm1}
\end{split}
\end{equation}
\vskip-15pt
\begin{equation}
\begin{split}
\hat\sigma_{gq}^{(2),\rm h} &=  \sigma_0
 \bigg\{\frac{11}{27} 
   + \frac{29}{6}\,\zeta_2 
   + \frac{311}{18}\,\zeta_3 
   + \frac{13}{81}\,n_f
 \\&\mbox{}
 + \lnxm{}\,\bigg[
   \frac{341}{18} 
   - \frac{50}{9}\,\zeta_2 
   - \frac{2}{3} n_f
 \bigg]
 \\&\mbox{}
 + \lnxm{2}\,\left( 
   \frac{85}{36} 
   + \frac{1}{18}\,n_f
 \right)
 + \frac{367}{54}\,\lnxm{3}
\\&
 + (1-x)\bigg[
 -\frac{959}{18} 
 + \frac{433}{9}\,\lnxm{} 
 - \frac{33}{2}\,\lnxm{2}
\\&
 + 8\,\zeta_2
 + \frac{4}{9}\,n_f\,\lnxm{}\bigg]
 + \ldots
 \bigg\}\,,
\label{eq::qgxm1}
\end{split}
\end{equation}
and
\begin{equation}
\begin{split}
&\hat\sigma^{(2),\rm h}_{q\bar q,\rm NS} = 
\hat\sigma^{(2),\rm h}_{q\bar q,\rm S} =
\hat\sigma^{(2),\rm h}_{qq,\rm NS} = \hat\sigma^{(2),\rm h}_{qq,\rm S} =
\\&\
 \sigma_0\bigg\{
 (1-x)\bigg[\frac{20}{9} 
- \frac{16}{9}\,\lnxm{} 
+ \frac{16}{9}\,\lnxm{2} 
- \frac{16}{9}\,\zeta_2\bigg]
+ \ldots\bigg\}\,.
\label{eq::qqxm1}
\end{split}
\end{equation}
For the sake of brevity, we have suppressed explicitly\break scale
dependent terms by setting $\mu_F = \mu_R = M_H$ (they can be readily
reconstructed using scale invariance) and displayed terms only to
order $(1-x)^1$.  Terms to order $(1-x)^1$ dominate the corrections
(see \fig{fig::nnloexp14}), but we include terms to order $(1-x)^{16}$
for all sub-processes in our numerical analysis.  The labels ``NS''
and ``S'' in \eqn{eq::qqxm1} denote the flavor non-singlet and singlet
quark contributions, respectively. The four contributions are equal
only to order $(1-x)^1$; their expansions differ at higher orders of
$(1-x)$ (except that $\hat\sigma^{(2),\rm h}_{q\bar q,\rm S} =
\hat\sigma^{(2),\rm h}_{qq,\rm S}$ exactly).  We note in passing that
our explicit calculation confirms the value for the coefficient
$c_{03}^{(2)}$ for the gluon-gluon subprocess derived in
\reference{Kramer:1998iq}.


\section{Hadronic Results}\label{sec::hadres}
The hadronic cross section $\sigma$ is related to the 
partonic\break cross section through a convolution with the parton
distribution functions.
It has been argued~\cite{Catani:2001cr} that convergence is improved
by pulling out a factor of $x$ from $\hat\sigma_{ij}$ before expanding
in $(1-x)$. We indeed observe a more stable behavior at low orders of
$(1-x)$ and will adopt\break this prescription in what follows.
Beyond fifth order, however, it is irrelevant which is used.

\begin{figure}[ht]
\includegraphics[width=\columnwidth]{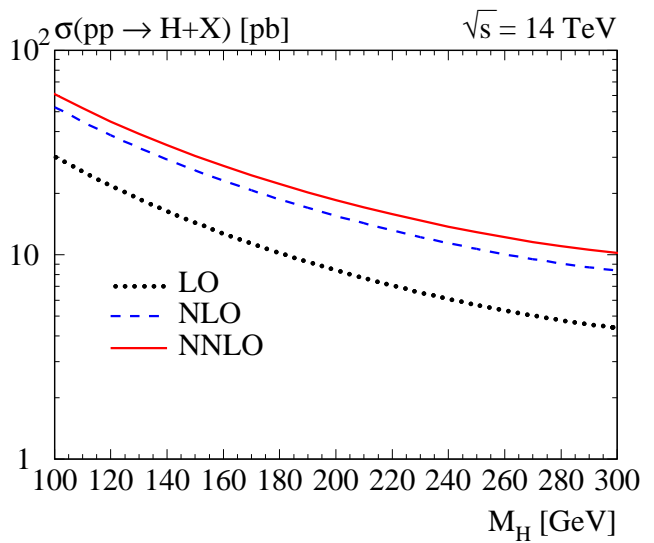}
      \caption[]{\label{fig::sigmannlo} {\abbrev LO} ({\it dotted}),
      {\abbrev NLO} ({\it dashed}) and {\abbrev NNLO} ({\it solid})
      cross sections for Higgs production at the {\abbrev LHC} ($\mu_F
      = \mu_R = \mhiggs$).  In each case, we weight the cross section
      by the ratio of the {\abbrev LO} cross section in the full
      theory ($M_t = 175$\,GeV) to the {\abbrev LO} cross section in
      the effective theory (\eqn{eq::efflag}).}
\end{figure}

In \fig{fig::sigmannlo}, we show the cross section at {\abbrev LO},
{\abbrev NLO} and {\abbrev NNLO}.  At each order, we use the
corresponding {\abbrev MRST} parton distribution set~\footnote{We thank
  R.~Roberts for providing us with the distributions and a draft of
  \reference{Martin:2002dr} prior to
  publication.}~\cite{Martin:2001es,Martin:2002dr}.  The {\abbrev NNLO}
distributions are based upon approximations of the three-loop
splitting functions~\cite{vanNeerven:2000wp}.  Studies using other
parton distributions, including the {\abbrev NNLO} distributions of
Alekhin~\cite{Alekhin:2001ih} will be presented elsewhere.

We next look at the quality of the expansion that we use for the
evaluation of the {\abbrev NNLO} corrections.
\begin{figure}[ht]
\includegraphics[width=\columnwidth]{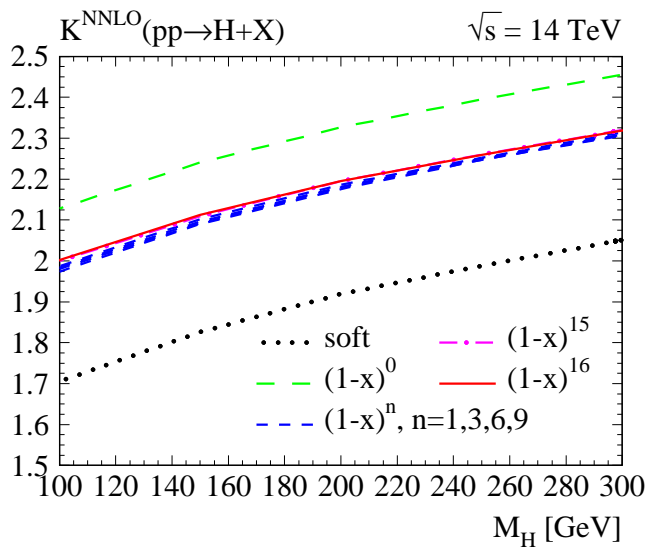}
      \caption[]{\label{fig::nnloexp14}
      $K$-factor for Higgs production at the {\abbrev LHC}.  Each line
      corresponds to a different order in the expansion in $(1-x)$. The
      renormalization and factorization scales are set to $\mhiggs$.}
\end{figure}
\fig{fig::nnloexp14} shows the {\abbrev NNLO} $K$-factor ($K^{\rm
NNLO}\equiv\sigma^{\rm NNLO}/\sigma^{\rm LO}$) for the {\abbrev LHC}
starting from the purely soft limit $\propto (1-x)^{-1}$ and adding
successively higher orders in the expansion in $(1-x)$ up to order
$(1-x)^{16}$.  Clearly, the convergence is very good: beyond order
$(1-x)^1$, the curves differ by less than $1\%$.  Observe that the
purely soft contribution underestimates the true result by about
10-15\%, while the next term in the expansion, $\propto (1-x)^0$,
overestimates it by about 5\%.  Note that the approximation up to
$(1-x)^0$\break is not the same as the ``soft+sl''-result of
\reference{Harlander:2001is} or the ``{\abbrev SVC}''-result of
\reference{Catani:2001ic}, since these include only the $\ln^3(1-x)$
terms at that order.

\begin{figure}[ht]
\includegraphics[width=\columnwidth]{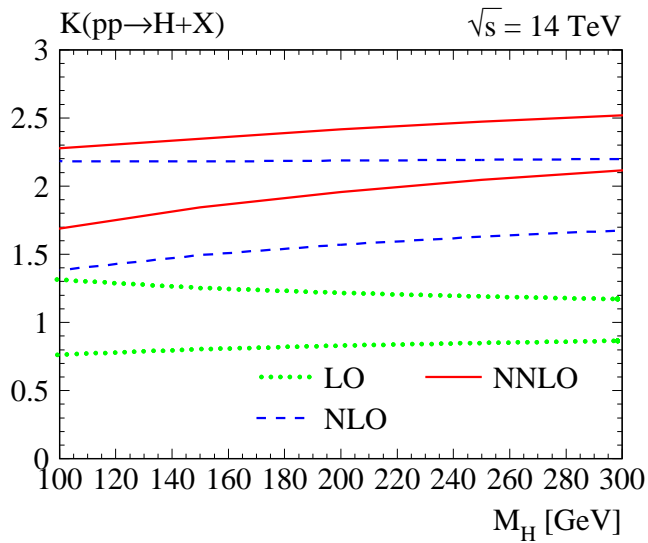}
      \caption[]{\label{fig::all14murf} Scale dependence at the
      {\abbrev LHC}. The lower curve of each pair corresponds to
      $\mu_R = 2\mhiggs$, $\mu_F=\mhiggs/2$, the upper to
      $\mu_R =\mhiggs/2$, $\mu_F=2\mhiggs$.  The $K$-factor is
      computed with respect to the {\abbrev LO} cross section at
      $\mu_R = \mu_F =\mhiggs$.}
\end{figure}

We next consider the renormalization scale ($\mu_R$) and factorization
scale ($\mu_F$) dependence of the $K$-factors. At the {\abbrev LHC},
we observe that the $\mu_F$ and $\mu_R$ dependence has the opposite
sign.  In order to arrive at a conservative estimate of the scale
dependence, we display two curves corresponding to the values
$(\mu_R,\mu_F) = (2\mhiggs,\mhiggs/2)$ and $(\mhiggs/2,2\mhiggs)$ (see
\fig{fig::all14murf}).
The scale dependence is reduced when going from {\abbrev NLO} to
{\abbrev NNLO} and, in contrast to the results in
\reference{Harlander:2001is}, the perturbative series up to {\abbrev
NNLO} appears to be well behaved. The reason is that both the
newly calculated contributions from hard radiation and the effect of
the previously unavailable set of {\abbrev NNLO} parton distribution
functions reduce the {\abbrev NNLO} cross section. Detailed studies of
the individual effects will be presented in a forthcoming
paper.

\begin{figure}[ht]
\includegraphics[width=\columnwidth]{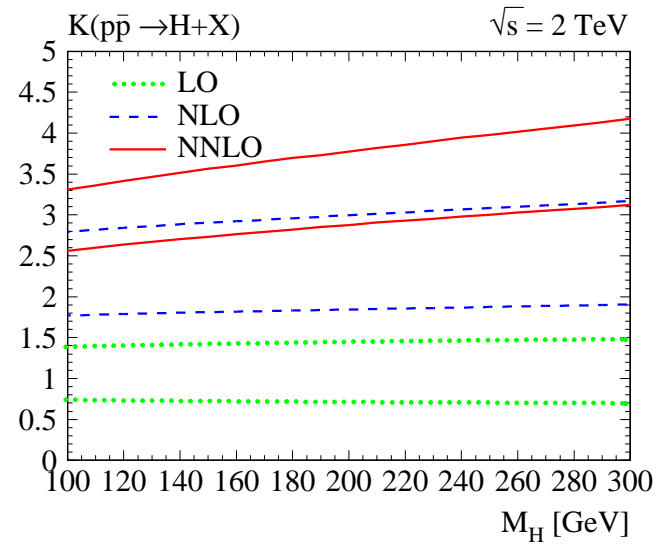}
      \caption[]{\label{fig::all2mu} Scale dependence for Tevatron
      Run~II. The lower curve of each pair corresponds to $\mu_R = \mu_F
      = 2\mhiggs$, the upper to $\mu_R = \mu_F = \mhiggs/2$.}
\end{figure}

Fig.~\ref{fig::all2mu} shows the results for the Tevatron at a
center-of-mass energy of $\sqrt{s}=2$\,TeV.
Here the dependence on $\mu_R$ and $\mu_F$ has the same sign, so we set
$\mu_R = \mu_F \equiv \mu$ and vary $\mu$ between $\mhiggs/2$ and
$2\mhiggs$. The $K$-factor is larger than for the {\abbrev LHC}, but
the perturbative convergence and the scale dependence are
satisfactory.


\section{Conclusions}
We have computed the {\abbrev NNLO} corrections to inclusive Higgs
production at hadron colliders.  We find reasonable perturbative
convergence and reduced scale dependence.

\mbox{}

\paragraph*{Acknowledgements:}
The work of R.V.H. was supported in part by Deutsche
Forschungsgemeinschaft.  The work of W.B.K. was supported by the
U.~S.~Department of Energy under Contract No.~DE-AC02-98CH10886.



\end{document}